\documentclass{phostproc}

\title{Effect of diffusion-induced element accumulation on the opacity inside B stars}
\author{Hui-Bon-Hoa, A., Vauclair, S.}

\affiliation{IRAP, Universit\'e de Toulouse, CNRS, UPS, CNES, Toulouse, France}

\shorttitle{Effect of diffusion on B stars inner opacity}
\shortauthors{Hui-Bon-Hoa A. \& Vauclair S.}
\abs{Stellar models with homogeneous abundances fail to reproduce the pulsation frequencies of early B-type stars. Their oscillations are excited by kappa-mechanism involving the Fe-peak elements where they are main contributors to the opacity (the "Z-bump") and a ad hoc increase of the opacity in these layers is necessary to match the observations.
We test whether atomic diffusion can induce such an opacity increase through Fe and Ni accumulations in the Z-bump. With models computed using the Toulouse–Geneva Evolution Code, we show that atomic diffusion changes the abundance profiles inside the star, leading to an overabundance of the iron-peak elements in the upper envelope. The opacity may reach the amount required by seismic studies, provided that fingering mixing, which extends the size of the overabundance zone, is taken into account. Mass-loss is also required to evolve the model until the end of the main sequence.}

\begin{document}

\maketitle

\section{Introduction}

Pulsations with acoustic modes ($p$-modes) and/or gravity modes ($g$-modes) are observed in a certain number of B-type stars. Some show only $p$-modes ($\beta$ Cep pulsators), some others only $g$-modes (Slowly Pulsating B stars, SPB), whereas some exhibit both types of modes (hybrid pulsators). The driving is due to the $\kappa$-mechanism involving the iron peak elements where they are main contributors to the opacity (the so-called "$Z$-bump"), around log~T=5.3. Stellar models computed with homogeneous abundances and currently available opacity datasets fail to reproduce the observed frequencies. Suitable matches require an increase of opacity in the driving region with ad hoc profiles \citep[e.g., ][]{2004MNRAS.350.1022P, 2017MNRAS.466.2284D}.\\
Efforts have been made by several authors to search for any missing opacity from the atomic physics point of view \citep[e.g., ][]{2016ApJ...823...78T}. Indeed, computed opacities may be underestimated compared to experimental measurements, as shown by \cite{2015Natur.517...56B} for iron at the bottom of the solar convective zone.\\
On another hand, an increase of opacity in the $Z$-bump can be due to a greater amount of the iron-group elements at this location \citep{2004MNRAS.350.1022P}. Local element accumulations can be created by atomic diffusion, which builds abundance stratifications of the chemicals when macroscopic mixing processes are weak enough. Each chemical species then migrates in the stellar medium, mainly under the effect of gravity, dragging the elements towards the stellar center, and radiative acceleration, which pushes them outwards. The latter depends on the ability of each species to absorb photons. As a consequence, individual elements may accumulate or be depleted in specific layers according to the variation of the radiative acceleration acting on them. Radiative accelerations and opacities are closely linked (e.g., their maximum will be at the same location) since both depend on the photon absorption properties of a given ion.

For B stars, a first study of the impact of diffusion has been performed by \cite{2006ASPC..349..201B}, with some simplifying assumptions in the computation of the radiative accelerations. Using a stellar evolution code in which the radiative accelerations are calculated consistently with the time variation of the abundances, \cite{2018A&A...610L..15H} (hereafter HV18) investigated how the opacities could be locally enhanced by an accumulation of Fe and Ni created by atomic diffusion, taking into account fingering mixing and mass-loss. Here, we extend their study with updated values of the Ni radiative accelerations, which are now adjusted to the Opacity Project (OP) data \citep{2005MNRAS.362L...1S}. Also, C, N, and O having significant mass fractions, they were considered in the diffusion computation to test their effect on the triggering of fingering mixing.\\
We first describe the main features of the code used in the calculations. Then we present our results and discuss them in the light of recent observations of the hybrid pulsating B star $\nu$ Eri, and conclude with ongoing and future work. 

\section{Method}

\subsection{Stellar models with atomic diffusion}
The stellar models are computed using an optimised version of the Toulouse-Geneva Evolution Code (TGEC, see HV18 for details), which includes atomic diffusion with radiative accelerations. The following isotopes are treated in details: H, $^3$He, $^4$He, $^6$Li, $^7$Li, $^9$Be, $^{10}$B,$^{12}$C, $^{13}$C, $^{14}$N, $^{15}$N, $^{16}$O, $^{17}$O, $^{18}$O, $^{20}$Ne, $^{22}$Ne, $^{24}$Mg, $^{25}$Mg, $^{26}$Mg, $^{40}$Ca, and $^{56}$Fe, to which we add nickel, since its contribution to the opacity is important in the oscillation driving layers.

Atomic diffusion is computed using the \cite{1970mtnu.book.....C} formalism, in which the chemicals move with respect to the dominant species, namely hydrogen.  We use the diffusion coefficients derived by \cite{1986ApJS...61..177P} and the OPAL2001 equation of state \citep{2002ApJ...576.1064R}. The nuclear reaction rates are from the NACRE compilation \citep{1999AIPC..495..365A}. At each time step of the evolution, the stellar structure is converged with Rosseland mean opacities computed consistently with the current abundance of each element at each layer of the star. We use the Opacity Project (OP) data and codes \citep{2005MNRAS.360..458B}, which are modified to enhance performance.
Dynamical convection zones are computed using the mixing length formalism with a mixing length parameter of 1.8. We assume that they are instantaneously homogenised. Fingering mixing is triggered when inversions of mean molecular weight occur (e.g., through the stratification of the elements) and is treated as a turbulent mixing for which we use the diffusion coefficient derived by \cite{2013ApJ...768...34B}.

\subsection{Computation of radiative accelerations}
Radiative accelerations are computed with a parametrised prescription, the Single-Valued Parameter (SVP) approximation \citep{2002MNRAS.332..891A,2004MNRAS.352.1329L}. By the use of analytical functions, this approach allows faster computations compared to the interpolation in precomputed tables of radiative accelerations or detailed calculations using monochromatic opacities. The agreement with those methods is very good since the coefficients of the SVP functions are set to adjust the detailed computations.
Still, the current set of SVP parameters is limited to stars of mass between 1 and 5~M$_{\odot}$. For Fe, HV18 remarked trends for these parameters with respect to the stellar mass, which allowed to extrapolate their values for higher masses using a power law function. We tried to apply the same method for C, N, and O but the extrapolations were not satisfactory. Instead, we fitted the total radiative accelerations with respect to the OP server values.

\begin{figure}
	\centering
	\includegraphics[width=1.1\linewidth]{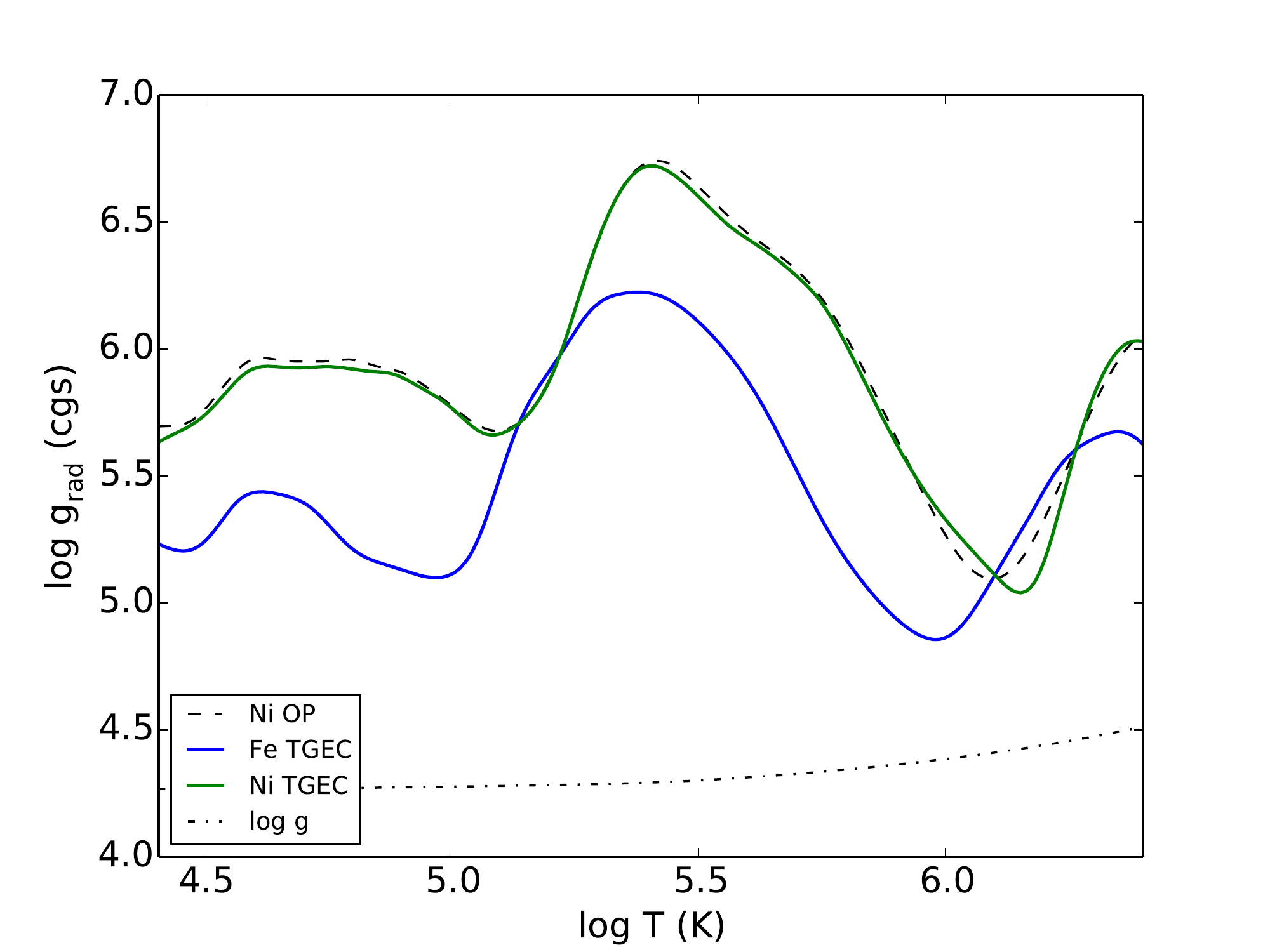}
	\caption{SVP radiative accelerations for Fe (blue solid line), and Ni (green solid line). For comparison, the dashed line shows the radiative accelerations for Ni from the OP server. The dash-dotted line denote the local gravity.}
	\label{g_rad}
\end{figure}

At present, nickel is not considered in the SVP tables so that we have to estimate all its SVP parameters from scratch.  As in HV18, we assume that the dependence of the radiative accelerations for each Ni ion with respect to its concentration is the same as for the Fe ion with the same number of electrons. We then fit the total Ni radiative accelerations to those of the OP server. Figure~\ref{g_rad} shows the behaviour of the radiative accelerations of Fe and Ni computed with the SVP method with respect to the local temperature (blue and green solid lines respectively) for a solar homogeneous abundance. The OP radiative acceleration profile for nickel (dashed line) is plotted to illustrate the good agreement between the two methods.\\
In HV18, the radiative accelerations for Ni were obtained through a fit with respect to the Montpellier-Montr\'eal code \citep{1998ApJ...504..539T}, which implements the OPAL opacities \citep[and references therein]{1996ApJ...464..943I}. In comparison, the present radiative acceleration profile for Ni shows a slightly stronger and shallower maximum in the $Z$-bump for our B star model.

\section{Results}
\begin{figure}
	\centering
	\includegraphics[width=1.1\linewidth]{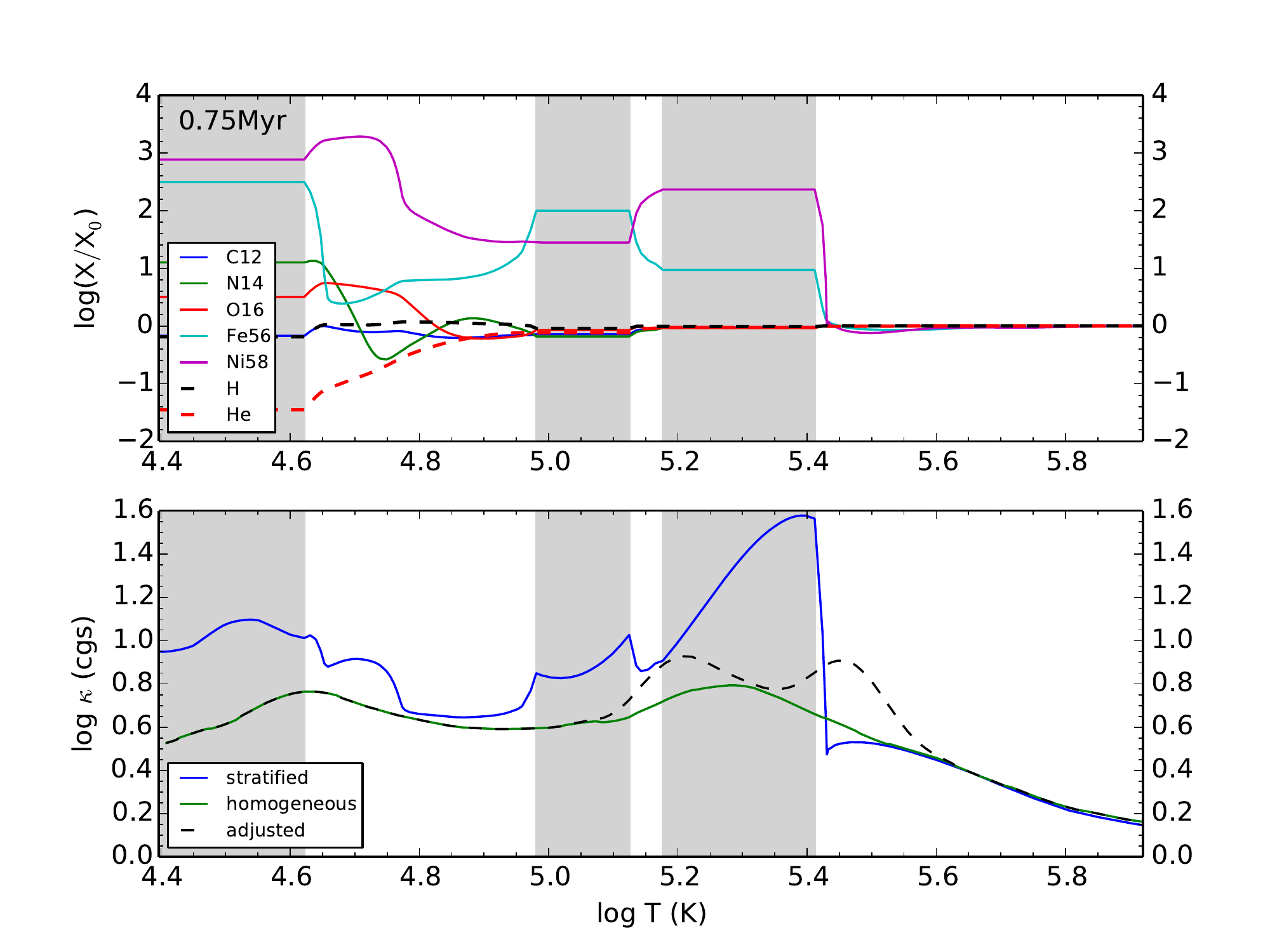}
	\caption{Model of a 9.5 M$_{\odot}$ star including only the effects of atomic diffusion.
Upper panel: mass fractions ratios with respect to initial mass abundance for C, N, O, Fe, Ni (blue, green, red, light blue, and purple solid lines, respectively), He (red dashed line), and H (black dashed line).
Lower panel: Rosseland opacity with homogeneous abundances (green solid line), and profile with stratified concentrations (blue solid line). The grey areas show the convective zones. The total opacity is increased where Fe and Ni accumulate. The dashed line shows the opacity modified according to asteroseismic observations, with the method and values of \cite{2017MNRAS.466.2284D}.}
	\label{diffNoTh}
\end{figure}

We computed a 9.5~M$_{\odot}$ model (the approximate mass of $\nu$ Eri) in which only He, C, N, O, Fe and, Ni were allowed to diffuse, the other elements keeping their initial (solar) abundance. The first four elements are considered owing to their significant contribution to the mean molecular weight and therefore their potential influence on the fingering mixing. The evolution is computed from the pre-Main Sequence (which lasts 0.22~Myr), and diffusion is set on at the beginning of the Main Sequence.\\
The first model was evolved assuming no fingering mixing and no mass-loss. Only atomic diffusion and convection are taken into account in the abundance evolution of the chemicals. Figure~\ref{diffNoTh} shows the abundance and opacity profiles at 0.75~Myr. In the upper panel, we can see the accumulations of Fe (light blue curve) around log~T=5.05, and that of Ni (purple curve) at log~T=5.3, and the corresponding increase of opacity (lower panel, blue solid line). The enhancement is significant in the Z-bump because the accumulation zones of Fe and Ni are broadened thanks to convection. Compared to the opacity profile constrained by asteroseismology, ours is quite different in shape and strength. Here the computations were stopped long before the end of the Main Sequence (which lasts \~24~Myr) to avoid too strong a surface abundance of iron.

\begin{figure}
	\centering
	\includegraphics[width=1.1\linewidth]{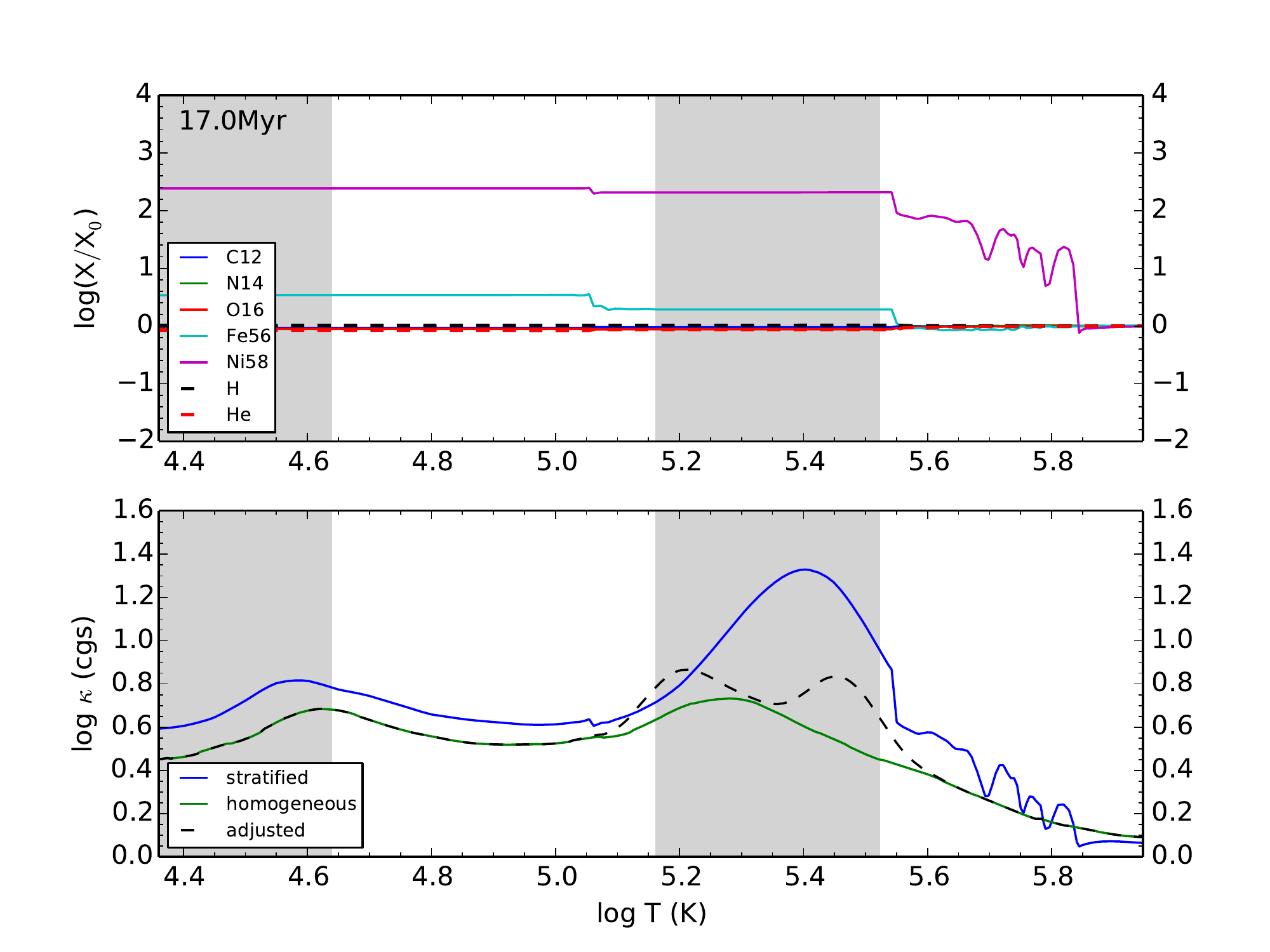}
	\caption{Same as Fig.~\ref{diffNoTh} but with fingering mixing and mass-loss. The abundance profiles of Fe and Ni are nearly flat due to fingering mixing.}
	\label{diffTh}
\end{figure}

In Fig.~\ref{diffTh}, the computations include fingering mixing and mass-loss with a rate of $\mathrm{10^{-12}~M_{\odot}/yr}$, typical for this kind of stars \citep{1989MNRAS.241..721P}. As a result, the amount of material expelled by mass-loss is great enough to avoid too strong surface abundances of iron and nickel, and allows to evolve the model up to the age of $\nu$ Eri \citep[around 17~Myr,][]{2004MNRAS.350.1022P}. The abundances of Fe and Ni growing much faster than the depletion of He, C, N, and O, inverse molecular weight gradients appear, leading to fingering mixing. The Fe and Ni accumulation zones are then broadened, leading to rather flat abundance profiles. The overabundance of Fe being lower than that of Ni, the opacity maximum is shifted towards the location of the maximum contribution of Ni (around log~T=5.4). Again, our opacity curve has a different shape compared to that computed by \cite{2017MNRAS.466.2284D}, with an unique broad peak extending from log~T=5.2 to log~T=5.6. In $\nu$~Eri, \cite{2012A&A...539A.143N} derived quasi-solar surface abundances for C, N, O and Fe. Our results are globally consistent with these observations, although we obtain a slightly overabundant iron compared to the solar value. Ni is strongly overabundant but has not been addressed by \cite{2012A&A...539A.143N}.

We now have to check if such an opacity profile can provide oscillation frequencies consistent with the observations. As \cite{2012A&A...539A.143N} derived solar values for the surface abundances of most of the oscillating B-type stars of their sample, we plan to extend our study to check whether this constraint can be satisfied with our models.

\section{Conclusion}
About fifty years ago, atomic diffusion was first invoked to account for abundance anomalies at the surface of Chemically Peculiar stars. With the advent of asteroseismic data of increasing sensitivity and precision, considering its effects on the inner structure is now mandatory. Here, we show that atomic diffusion, along with the fingering mixing that it naturally triggers, can lead to a strong opacity enhancement in the layers where the oscillations of B-type stars are driven. Further computations are needed to check whether the observed pulsation frequencies can be excited using our models. The surface abundances put additional constraints that our models have to satisfy. 

%\section*{Acknowledgments}

\bibliographystyle{phostproc}
\bibliography{diffusionBstars}

\end{document}